\documentclass[11pt]{article}
\usepackage[a4paper,hmarginratio=1:1,vmarginratio=2:3,totalwidth=16.25cm,
totalheight=21.6cm]{geometry}  
   
  \usepackage{latexsym}
\usepackage{epsfig}

\newcommand{\Li}{\mbox{Li}}

\newcommand{\ra}{\rightarrow}
\newcommand{\be}{\begin{equation}}
\newcommand{\ee}{\end{equation}}
\newcommand{\bea}{\begin{eqnarray}}
\newcommand{\eea}{\end{eqnarray}}

\DeclareMathSymbol{\mg}{\mathrel}{symbols}{"1D}

\newcommand{\cJ}{{\cal J}}
\newcommand{\cO}{{\cal O}}

\newcommand{\cR}{{\cal R}}

\newcommand{\cK}{{\cal K}}
\newcommand{\cL}{{\cal L}}
\newcommand{\cB}{{\cal B}}
\newcommand{\cF}{{\cal F}}
\newcommand{\cH}{{\cal H}}

\newcommand{\bZ}{{\bf Z}}
\newcommand{\tabu}[2]{\begin{tabular}{#1} #2 \end{tabular}}
\newcounter{oldcounter} 
\addtocounter{equation}{1}
\begin{document} 
\begin{flushright}  
{DAMTP-2004-78}\\
\end{flushright}  
\vskip 3 cm
\begin{center} 
{\Large 
{\bf  Higher Derivative  Operators as loop counterterms  \\
\vspace{0.2cm}
in one-dimensional field theory orbifolds.\\}}
\vspace{1.9cm} 
{\bf D.M. Ghilencea}\\
\vspace{1.cm} 
{\it D.A.M.T.P., Centre for Mathematical Sciences, 
University of Cambridge, \\
Wilberforce Road, Cambridge CB3 OWA, United Kingdom}\\
\end{center} 
\vspace{2cm}
\begin{abstract}
\noindent
Using  a 5D N=1 supersymmetric  toy-model compactified on 
$S_1/(Z_2\times Z_2')$,  with a ``brane-localised'' superpotential, 
it is  shown that higher  (dimension) derivative operators are
generated as one-loop counterterms  to the (mass)$^2$ of the  zero-mode 
scalar field, to ensure the quantum consistency of the model. Such  
operators are just a result of the compactification and integration of 
the bulk modes.  They are relevant for  the UV momentum scale dependence of 
the  (mass)$^2$ of the zero-mode scalar field, regarded as a Higgs
field in more realistic models.
While suppressed for a small  compactification radius R, these
operators  can affect the predictive power of models  with a large 
value for R. A general method is also  provided  for a careful evaluation 
of infinite sums of  4D divergent   loop-integrals  (of  Feynman
diagrams)  present in field theory orbifolds. With minimal changes, 
this method  can  be applied to specific orbifold models for a simple 
evaluation of their  radiative corrections and the overall divergences. 
\end{abstract}
\newpage

\section{Introduction}
The physics of extra dimensions received a renewed
interest in the context of field theory orbifolds which attempt to
derive  SM-like  models  (or supersymmetric versions)
from compactifications 
of a higher dimensional theory (see for example 
refs.\cite{dpq}-\cite{mr}). In this work we consider a rather 
simple  5D model compactified on $S_1/(Z_2\!\times\! Z_2')$ 
to illustrate further quantum effects, so far neglected.

The motivations of this study are described in the following.
In the context of an orbifold compactification of a 5D model, a
 natural question to ask is whether higher dimension operators 
are important at one loop (assuming they were not introduced in 
the tree level action). The study of higher dimensional 
operators is important for they  can affect the predictive power of 
the model by introducing  additional parameters in the theory.
This may be particularly relevant in models with a low
 compactification scale.
For  renormalisable theories,  if they are  not added at the tree
level, they will not be generated as counterterms by radiative 
corrections. However, in higher
dimensional (non-renormalisable) models such as that  considered here 
this is not  true.  In some cases higher dimensional
 (derivative) operators are  radiatively generated as one-loop
 counterterms to the couplings or masses of the model,  
see for example the case of gauge couplings 
\cite{Oliver:2003cy}, \cite{Ghilencea:2003xj} in 
two-dimensional compactifications.  While their presence there was 
related to the two-dimensional nature of the compactification, we
shall see  that the same technical reason which generated them at 
one-loop\footnote{This was due to the levels  degeneracy and the
 presence of two Kaluza-Klein 
sums acting on the loop integral~\cite{Ghilencea:2003xj}.}
is at work even in the case of simpler, one-dimensional orbifolds.

A second motivation is to provide in an orbifold  model  an 
Wilsonian picture  for the dependence on the high (UV) scale
of the radiative corrections (induced by the Kaluza-Klein (KK) modes)
to the couplings or masses of the theory. 
This is done in a way which  keeps manifest symmetries generic in  this 
class of models (gauge, etc) (for a   discussion of 
symmetries, anomalies and regularisation issues in field theory orbifolds  
see \cite{GrootNibbelink:2002qp}-\cite{Ghilencea:2001bv}). 
The technique we employ  can be applied whenever one must sum 
infinitely many KK contributions from 4D  divergent Feynman diagrams
and  only requires a DR regulator $\epsilon$ ($\epsilon\ra 0$) 
for the associated  loop-integrals.
To find the  scale dependence of the radiative corrections 
in the  DR scheme one must evaluate the relevant loop diagrams for
 non-vanishing external momentum  ($q^2$).  Usually one takes
$q^2=0$ since the external momentum is much smaller than 
the mass  of the KK modes in the loop. In general this is appropriate
for  a high compactification scale, but in this work we  keep $1/R$
arbitrary. Moreover, at $q^2\!=\!0$ additional would-be 
divergences of radiative corrections
 (such as for example 
$q^{2\,n} R^{2 n -2}/\epsilon\,$ for a one-loop corrected 
(mass)$^2$), induced by   multiple sums over the KK modes, 
cannot be ``seen''\footnote{Such divergences can be due to higher
derivative operators, see discussion below.}. 
With these remarks,  the momentum scale $q^2$ is regarded as 
the physical  scale of the  Wilsonian picture of the theory.
As an example, we shall consider a brane-localised superpotential in a 
minimal 5D N=1 supersymmetric model compactified on 
$S^1/(Z_2\!\times \! Z_2')$ and consider the one-loop effect on 
the (mass)$^2$ of a
KK  zero-mode scalar field (hereafter denoted $\phi_{H,0}$). 
We thus address the dependence of
this (mass)$^2$ on the momentum scale $q^2$ to gain  information on  
the UV regime of the model. Our findings are relevant for
field theory orbifold models with similar  compactifications
\cite{dpq}, \cite{DiClemente:2001ge}-\cite{Arkani-Hamed:2001mi},
where our zero-mode scalar field $\phi_{H,0}$ 
is usually identified with a  Higgs field. In such models 
  $m_{\phi_{H,0}}^2(0)\!\sim\! 1/R^2$, 
  with implications for phenomenology \cite{dpq}-\cite{mr}.
Here we  compute $m_{\phi_{H,0}}^2(q^2)$
and its dependence under the (UV) scaling  of $q^2$.
Such dependence  cannot  be supplied  by  string  or
 field theory calculations of the loop expansion of the
corresponding  scalar
potential (derived at $q^2\!=\!0$ of the external legs) with tree
level   (fixed) values  of the couplings.

Another motivation is to clarify the {\it exact} link between 
higher  derivative operators (HDO) on orbifolds and the 5D nature 
of the initial theory.  As we shall see shortly, it turns out that 
the presence of HDO is strictly related to the {\it infinite} Kaluza-Klein 
sums associated  with the initial 5D theory. Therefore, the existence
of HDO  is just an effect of compactification of a higher
dimensional theory which  - although supersymmetric - is nevertheless
non-renormalisable,  and this  is an interesting finding. 
It  is useful to  note  the connection  of such aspects of compactification 
to dimensional crossover, coarse graining   and non-perturbative effects,
largely studied in the context of  Ising-like systems~\cite{O'Connor:zj}.
Such a  point of view can provide a rich insight into an  orbifold 
compactification,  the dependence  of the couplings or 
masses  on $q^2$ and the decoupling of {\it infinitely} 
many states, while varying $q^2$ relative to $1/R^2$.

Finally we  outline a  generalisation of the 
method used in the example below
for computing radiative corrections in field theory  orbifolds.  
The method  carefully  evaluates   the 
infinite  sums of  divergent loop integrals of the corrections, 
applies for  a very  general Kaluza-Klein mass spectrum 
and is presented in  Appendix \ref{appc}. We hope this will 
help  the careful study of  more realistic models.

\section{Higher derivative operators as  one-loop counterterms.}

To illustrate these
ideas  we consider the one-loop corrected
(mass)$^2$ of a (zero-mode) scalar field, 
in a 5D N=1 supersymmetric toy-model with 
hypermultiplet fields compactified on $S_1/(Z_2\!\times\! Z_2')$. 
Here $S_1$ has radius $R$ and we have the  identification 
$x_5\ra x_5+2\pi R$.
Under $\, Z_2: x_5\ra -x_5$,  while under $\, Z_2': x_5\ra -x_5 +\pi R$.
An N=2 hypermultiplet  decomposes into two four-dimensional N=1 
chiral superfields $M$ and $M^c$  of components 
$M=(\phi_M, \psi_M)$ and $M^c=(\phi_M^c,\psi_M^c)$, with
opposite quantum numbers. We also assume the index $M$ runs over the  
superfields  $Q, U, D, L, E$.
Their parities are fixed such as their (component) fermion fields are 
massless zero-modes fields, as one would like in a realistic
model aimed at reproducing the  low-energy SM ``massless'' 
 fermionic spectrum. This is made possible with the parities assignment 
in the table below.  The analysis  can  be extended to three 
families of fields, as any realistic model would require.  
Further, an  extra hypermultiplet is introduced, 
of N=1 components  $H=(\phi_H, \psi_H)$ and  $H^c=(\phi_H^c,\psi_H^c)$ 
whose  $Z_2'$ parity 
is chosen opposite   to that of the fields $M$. Such (allowed) choice 
ensures that the massless zero-mode is now a scalar rather than a
fermion. In a more  detailed model \cite{Barbieri:2000vh} this scalar may  be
identified  with the usual Higgs boson of the SM.
This different parity choice under $Z_2'$ distinguishes between the 
two types of hypermultiplets  on  $S_1/(Z_2\!\times\! Z_2')$. 
The  $(Z_2,Z_2')$ parities and the KK 
 masses are then 
\begin{center}\label{table}
\qquad
\tabu{ c | c | c | c | c | c | c |  c | c  }{4D fields  &  
$\phi_{M,n}$  &  $\phi_{M,n}^c$  &  $\psi_{M,n}$  & $\psi_{M,n}^c$ &    
$\phi_{H,n}$  &  $\phi_{H,n}^c$ & $\psi_{H,n}$  & $\psi_{H,n}^c$ \\ 
\hline 
$(Z_2, Z_2')$  &  $+,-$ &  $-,+$ &  $+,+$ & $-,-$   &
$+,+$  & $-,-$ & $+,-$  & $-,+$  \\ 
modes $n$  & $\geq 0$  & $\geq 0$  & $\geq 0$  & $\geq 1$  & $\geq 0$
& $\geq 1$  & $\geq 0$  & $\geq 0$\\
(mass)$\times R$ & $2n\!+\!\!1$ & $2n\!+\!\!1$ & $2n$  & $2n$ & $2n$ &  $2n$ & 
$2n\!+\!\!1$ & $2n\!+\!\!1$
\\
}
\\
\end{center}
The KK  fields come with the wavefunctions
\begin{eqnarray}
\phi_{M,n}:\, \cos\frac{(2n+1) x_5}{R}, \qquad
\phi_{M,n}^c:\, \sin\frac{(2n+1) x_5}{R}, \qquad
\psi_{M,n}: \,\cos\frac{2 n x_5}{R},\qquad
\psi_{M,n}^c:\, \sin \frac{2 n x_5}{R}
\nonumber\\
\psi_{H,n}: \,\cos \frac{(2n+1) x_5}{R},\qquad
\psi_{H,n}^c: \,\sin \frac{(2n+1) x_5}{R},\qquad
\phi_{H,n}: \,\cos \frac{2 n x_5}{R},\qquad
\phi_{H,n}^c:\, \sin \frac{2n x_5}{R}\,
\end{eqnarray}
The next step would be to introduce 
vector multiplets and the corresponding gauge 
interactions\footnote{If included,  with
this hypermultiplet content,  a quadratic divergence 
exists due to Fayet-Iliopoulos term~\cite{Ghilencea:2001bw}.}.
However, we will not do so, and restrict the spectrum to the
matter content given above.
Although  this assumption is  not phenomenologically viable, it 
is  appropriate to illustrate the main point of 
this work.  Further, the interaction that we consider is
\begin{equation}\label{int}
L=\int d x_5 \, \frac{1}{2}\Big[\delta(x_5)+\delta(x_5-\pi R)\Big] 
\int d^2 \theta\,\, \lambda\,\, Q \, U\, H +h.c.
\\
\nonumber
\end{equation} 
Note that {\it all fields} $Q, U, H$ in (\ref{int}) are bulk fields 
(have associated KK modes). Here $\lambda\sim \textrm{(mass)}^{-3/2}$ and the
fields have dimension (mass)$^{3/2}$. This minimal interaction is 
very  common  in  5D extensions of  the SM
(see for  example \cite{Barbieri:2000vh},  \cite{Arkani-Hamed:2001mi}),
with important implications for 
phenomenology\footnote{This minimal model may also  be recovered in
  the formal limit 
of vanishing gauge couplings  of the model in \cite{Barbieri:2000vh}.},
and this motivates its further analysis.
One may include additional Yukawa interactions at the other fixed 
points of the orbifold ($\pm \pi R/2$). Such interactions
can involve fields $Q, D, H$  or $L, E, H$ 
(with $\phi_M\!\ra\! \phi_M^{c \dagger}$).
A detailed analysis should also  consider overlapping
effects of the two types of fixed points and associated
interactions,  since only then would the ``global'' structure of
$S_1/(Z_2\!\times\! Z_2')$ be seen. This  involves however higher 
loop calculations and is beyond the 
purpose of this paper. For simplicity  the discussion below is 
restricted to the interaction  given in~(\ref{int}).

Our purpose is to compute the dependence of the
 one-loop correction to the (mass)$^2$ of the
zero-mode scalar $\phi_{H,0}$ of $H$, 
on the   (Euclidean) external momentum $q^2$ in the corresponding 
Feynman diagrams induced by interaction (\ref{int}).   
Previous  calculations of one-loop corrected (mass)$^2$ 
in 5D orbifold models were restricted to  $q^2=0$  when they  
simplify  considerably. Further, it is usually thought that,
to investigate the leading UV behaviour of the scalar field
mass in 5D orbifolds, it is  sufficient to consider
$m_{\phi_{H,0}}(0)$. However, in general in a 
{\it non-renormalisable} model  with infinitely many states 
in the loop, this picture is incomplete and new effects
emerge, as we shall see shortly.

After expanding interaction (\ref{int}) in component 
fields\footnote{See  ref.\cite{WB}. For interaction (\ref{int})
in full  component form  (onshell) see for 
example \cite{Barbieri:2000vh},  also \cite{Arkani-Hamed:2001mi}.} 
and  Fourier (Kaluza-Klein)  modes corresponding to $x_5$, 
one has three one-loop  Feynman diagrams for the 
two-point Green function with the zero-mode scalar~$\phi_{H,0}$
as external legs. The first has two vertices,  with
both $\phi_{Q,m}^c$ and  $\phi_{U,k}$ 
in the loop, coupling together to $\phi_{H,0}$ (and a similar contribution
 with $Q\!\leftrightarrow\!  U$). In addition to this 
there is a one-vertex 
diagram with only  $\phi_{Q,k}$ in the loop, which couples
to $\phi_{H,0}$ (a similar contribution with $\phi_{U,k}$).
For the fermionic contribution there is a
diagram  with both $\psi_{U,k}$ and $\psi_{Q,m}$  in the 
loop, coupling together  to $\phi_{H,0}$.
After evaluating these one-loop contributions in the DR scheme, one
finds (up to an overall colour factor, not written explicitly) 
the following formulae for the  
bosonic ($\cB$)  and  fermionic ($\cF$)  effects to the 
mass $m_{\phi_{H,0}}^2$  of the zero-mode scalar~$\phi_{H,0}$

\begin{eqnarray}\label{gre}
- i\, m_{\phi_{H,0}}^2(q^2)\bigg\vert_\cB
 & = & - i\, f_t^2 \sum_{k, m\geq 0}
\eta_{k,\phi}^2 \eta_{m,F}^2  \,\mu^{4-d}
\int
\frac{d^d p}{(2\pi)^d}
\frac{(p+q)^2}{(p^2+m^2_{\phi_{U,k}})\,((p+q)^2+m^2_{\phi_{Q,m}^c})}
\nonumber\\
\nonumber\\
- i\, m_{\phi_{H,0}}^2(q^2)\bigg\vert_\cF
 & =  & \,\, \,\,i \,f_t^2\sum_{k, m\geq 0} 
\eta_{k,\psi}^2 \eta_{m,\psi}^2 \, \mu^{4-d}
\int
\frac{d^d p}{(2\pi)^d}
\frac{p.(p+q)}{(p^2+m^2_{\psi_{U,k}})\,((p+q)^2+m^2_{\psi_{Q,m}})}
\\
\nonumber
\end{eqnarray}
Here $d=4-\epsilon$, with  $\epsilon\!\ra \! 0$ the DR regulator 
of  the momentum integrals, and $\mu$ is the usual finite, non-zero mass
scale that dimension regularisation introduces. $f_t$ is the 4D
coupling,  related to $\lambda$ by $\lambda=(\pi R)^{3/2} f_t$.
Further,
 $\eta^2_{k,\phi}=\eta^2_{m,F}=1$  for the bosonic sector and 
$\eta^2_{0,\psi}=1/2,$   $\eta^2_{k,\psi}=1$, ($k\geq 1$) for
fermions. These are wavefunction coefficients which take
account of different normalisation of the zero-mode fields.
 From the  table of parities  one has
\begin{eqnarray}\label{mss}
m_{\phi^c_Q,k} \!\!=\! \!  m_{\phi_{U},k}
\,=\,\frac{2}{R}\bigg[k+\frac{1}{2}\bigg],\,\, k\geq 0;\qquad 
\textrm{and:} \qquad
m_{\psi_U,k} \!\! =\!\!  m_{\psi_Q,k}\,=\,
\frac{2}{R}\, k,\,\,\, k\geq 0.
\end{eqnarray}
This  mass spectrum  is special in that it
allows one to re-write the one-loop correction in  
(\ref{gre}) as sums over the
whole set $\bZ$ of integers\footnote{
We use $\sum_{k_1\geq 0} h\big((k_1\!+\! 1/2)^2\big)\!=\!
\frac{1}{2}\sum_{k_1\in\bZ} h\big((k_1\!+\! 1/2)^2\big)$
and $\sum_{k_1\geq 0} \frac{1}{2^{\delta_{k_1,0}}} 
h\big(k_1^2\big)\!= \! \frac{1}{2}\sum_{k_1\in\bZ} h\big(k_1^2\big)$, 
$h$ is an arbitrary function; this step is possible 
if Kaluza-Klein masses have special values (\ref{mss}),
but not for those in (\ref{mass}).} 
and this simplifies our calculation considerably. 
Appendix~\ref{appc} extends  the calculation of (\ref{gre}) 
to  a more general  KK mass spectrum than that in (\ref{mss}), 
(which takes account 
of potential mass shifts induced by levels mixing if present),  
and with  important changes for the overall 
divergences of $m_{\phi_{H,0}}$. 
To see the difference, compare eq.(\ref{gre}) with 
(\ref{mss}) against eqs.(\ref{klgh}), (\ref{thr}) with (\ref{mass}) 
in Appendix~\ref{appc}.
Further, from  eqs.(\ref{gre}), (\ref{mss}) and Appendix \ref{appa} one has

\begin{eqnarray}\label{b}
-  m_{\phi_{H,0}}^2(q^2)\bigg\vert_\cB\!\!\!\!\!
& = &\!    \frac{-1}{4}\,  f_t^2 \,\mu^\epsilon\!\!\!
\sum_{k_1, k_2\in\bZ}
\int \frac{d^d p}{(2\pi)^d}
\frac{(p+q)^2}{((p+q)^2+ 4\, (k_2+1/2)^2/R^2)\,(p^2+4 \,(k_1+1/2)^2/R^2)}
\nonumber\\
\nonumber\\
\! &=&\!
\frac{- 1}{4}\, f_t^2 \frac{\kappa_\epsilon}{(4\pi R)^2} 
\int_0^1 dx\,
\bigg[ \frac{2-\epsilon/2}{\pi}\, \cJ_2[ 1/2,1/2,c]
+ q^2 R^2 (1-x)^2  \, \cJ_1[ 1/2,1/2,c]
\bigg]\qquad
\end{eqnarray}
and 
\begin{eqnarray}\label{f}
-  m_{\phi_{H,0}}^2(q^2)\bigg\vert_\cF\!\!\!\!\!
 & = & \! \;\frac{1}{4}  \, f_t^2\, \mu^\epsilon\!\!\!
\sum_{k_1,k_2\in\bZ}
\int
\frac{d^d p}{(2\pi)^d}
\frac{p.(p+q)}{((p+q)^2+4\, k_2^2/R^2)\,(p^2+4\, k_1^2/R^2)}
\nonumber\\
\nonumber\\
\! &=& \!\;
\frac{1}{4} \, f_t^2\, \frac{\kappa_\epsilon}{(4\pi R)^2} 
\int_0^1  dx \,
\bigg[ \frac{2-\epsilon/2}{\pi} \, \cJ_2[0,0, c]
+ q^2 R^2 x (x-1)  \, \cJ_1[ 0,0, c]
\bigg]\qquad\qquad\qquad
\\
\nonumber
\end{eqnarray}
with the notations $\kappa_\epsilon\equiv (2\pi R\,\mu)^\epsilon$
and 
\begin{eqnarray}\label{jjjs}
&&\cJ_j[ c_1,c_2,c]\equiv \sum_{k_1,k_2\in \bZ}\int_0^\infty
\frac{dt}{t^{j-\epsilon/2}}\, e^{-\pi\,
    t\,(c+a_1(k_1+c_1)^2+a_2(n_2+c_2)^2)},\quad j=1,2;\quad a_{1,2}, c>0; 
\nonumber\\
\nonumber\\
&& a_1 \equiv 4 \,(1-x), 
\quad a_2 \equiv 4\, x,  
\quad c\equiv x\,(1-x)\, q^2 R^2
\end{eqnarray}
The functions $\cJ_{1,2}$ are evaluated in Appendix \ref{J1andJ2}. 
The total one-loop correction to the mass $m^2_{\phi_{H,0}}(q^2)$
 is the sum of the two contributions in (\ref{b}), (\ref{f}). 
Note that only $\cJ_2$ contributes to
$m^2_{\phi_{H,0}}(0)$, and this may be seen by formally setting
$q^2=0$ in the second line of (\ref{b}) and (\ref{f}).

The quantities $\cJ_{1}$ and $\cJ_2$ have each a divergent and
a finite part. Let us first discuss  their divergent part 
 which contributes to $m_{\phi_{H,0}}$. Even though in $\cJ_{1,2}$ one
sums over the whole set $\bZ$ of integers, and no  finite set of levels
(i.e. modes) is excluded from their infinite sums, 
 their  integrals are still UV divergent (i.e. at $t\ra 0$). 
One shows that $\cJ_1$ and $\cJ_2$ have the behaviour 
\begin{eqnarray}\label{j1j2}
\cJ_j[c_1,c_2,c] =  \bigg[\frac{2}{\epsilon}\bigg]\,
 \frac{\big(-\pi  c\,\big)^j}{j\, \sqrt{a_1 a_2}}+\cO(\epsilon^0),\qquad
 j=1, 2.
% \\
% \nonumber
\end{eqnarray}
Although the  divergences in  $\cJ_1$ and $\cJ_2$ 
depend on $c\sim q^2 R^2$, they are  independent of the 
coefficients $c_1$ and $c_2$ which here are fixed to $0$ and $1/2$ 
by the initial orbifold parity conditions  on the 5D fields. 
Note the divergences of $\cJ_{1,2}$ are
only ``seen'' for  non-zero\footnote{If either $c_{1,2}$ are
non-integers  $\cJ_{1,2}$ are well defined even for $c=0$;
  then the  divergences of $\cJ_{1,2}$ are not ``seen''.} $c\sim
q^2 R^2$.

Let us  consider now the finite part of $\cJ_{1,2}$. For simplicity we 
 consider the case $c\!\ll\! 1$  respected under the 
 sufficient assumption $q^2\!\ll\! 1/R^2$.  Eq.(\ref{eqb4}) in 
Appendix~\ref{J1andJ2} gives, ignoring $\cO(c)$ terms 

\begin{eqnarray}\label{j1finite}
\cJ_1[0,0,c\ll 1]
& =& 
\frac{\pi c}{\sqrt{a_1 a_2}}
\bigg[\frac{-2}{\epsilon}\bigg]
-\ln \bigg[4 \pi e^{-\gamma} \big\vert
 \eta (i\, u)\big\vert^4 \frac{1}{a_2}\bigg]
-\ln (\pi e^\gamma c)
\nonumber\\
\nonumber\\
\cJ_1[\frac{1}{2},\frac{1}{2}, c\ll 1]& = & 
\frac{\pi c}{\sqrt{a_1 a_2}}
\bigg[\frac{-2}{\epsilon}\bigg]
-\ln  \bigg\vert \frac{\vartheta_1 
(1/2 -i u/2\vert i u)}{\eta(i \,u)}\,\,e^{-\pi\, u/4} \bigg \vert^2,
\qquad u\equiv \bigg[\frac{a_1}{a_2}\bigg]^\frac{1}{2}
\\
\nonumber
\end{eqnarray}

\noindent
In a similar way one shows (see eq.(\ref{jj1jj2}))

\begin{eqnarray}\label{j2finite}
\cJ_2[0,0,c\!\ll \!1] \!\!\!\!& = &\!\!\!\!\!
\frac{-\pi^2 c^2}{2 \sqrt{a_1 a_2}}\bigg[\frac{-2}{\epsilon}\bigg]
\!+\! \frac{\pi^2 a_1 u}{45}
\!+\! \frac{a_2}{\pi} \sum_{n\in\bZ}\!\bigg[ 
\Li_3(e^{-2 \pi\, u\,\vert n \vert})
\!+\! 2\pi u  \vert n\vert \,
 \Li_2(e^{-2\pi u \vert  n\vert})\bigg]\qquad\qquad\quad
\end{eqnarray}
\begin{eqnarray}
%\nonumber\\
\cJ_2 [\frac{1}{2},\frac{1}{2},c\!\ll \! 1 ] \!\!\!\! & = & 
\!\!\!\!\!
\frac{-\pi^2 c^2}{2 \sqrt{a_1 a_2}}\bigg[\frac{-2}{\epsilon}\bigg]
\! -\! \frac{7 \pi^2 a_1 u}{360}  
\! +\!
\frac{a_2}{\pi}\!\sum_{n\in\bZ} 
\bigg[\Li_3(-e^{-2\pi u \,\vert n\! +\!\frac{1}{2}\vert})
\! + \! 2\pi u  \Big\vert n\! +\!\! \frac{1}{2} \Big\vert
\, \Li_2(-e^{-2\pi \,u \vert n+\frac{1}{2}\vert})\!\bigg]
\nonumber
\\
\nonumber
\end{eqnarray}
where again  $\cO(c)$ terms were neglected. For results
without the restriction $c\ll 1$ see Appendix~\ref{J1andJ2}.
The divergences in $\cJ_1$ and  $\cJ_2$ in the last two sets of 
equations are
in agreement with the general case of (\ref{j1j2}) which does not
have the restriction $c\ll 1$.
Eqs. (\ref{b}), (\ref{f}) give the total contribution

\begin{eqnarray}\label{div}
-m_{\phi_{H,0}}^2 (q^2) & = & 
\frac{f_t^2}{32 \pi^3 R^2}\int_0^1 dx\, 
\Big[ \cJ_2[0,0,c]-  \cJ_2[{1}/{2},{1}/{2},c] \Big]
\nonumber\\
\nonumber\\
&+& \frac{f_t^2}{64\, \pi^2 R^2} 
\,(q^2 R^2) \int_0^1  \! dx  \Big[
x(x - 1) \cJ_1[0,0,c]
-(1-\!x)^2 \cJ_1[{1}/{2},{1}/{2},c] \Big]
\\
 \nonumber
\end{eqnarray}
Note the presence of the factor $q^2 R^2$ in front of the last
integral. Therefore, if one studies only $m_{\phi_h,0}^2(0)$
the second line in (\ref{div}) is absent. Further,
if  $c\sim q^2 R^2\!\ll\! 1$ the finite contributions from each  $\cJ_1$ are 
suppressed, but  the first term (divergent)
 in each $\cJ_1$ in (\ref{j1finite}), must be included  in (\ref{div}).
Alternatively, one can use $\cJ_1$ of eq.(\ref{j1finite})
and with $\cJ_2$ to order $\cO(c^2)$ derived from 
Appendix \ref{J1andJ2} or more generally, 
the  full expressions given there.

The divergence of $\cJ_2$ in (\ref{j2finite})
or more generally (\ref{j1j2}) cancels out  in $m^2_{\phi_{H, 0}}$
because  $\cJ_2[0,0,c]$ and  $\cJ_2[1/2,1/2,c]$ have equal 
coefficient in eq.(\ref{div}). This cancellation is
ensured by the equal number of bosonic and fermionic 
degrees of freedom, a remnant of initial 5D supersymmetry;
in this $c^2\sim q^4 R^4$ plays a regularisation role. 
For $q^2\!\ra\! 0$ the one-loop finite contribution $m_{\phi_H,0}^2(0)$ 
is due to  $\cJ_2$ only and the result agrees numerically with that in
\cite{Barbieri:2000vh} for a vanishing gauge coupling. 
More generally,  from eq.(\ref{div})
\begin{eqnarray}\label{dr}
m_{\phi_{H, 0}}^2(q^2)=m_{\phi_{H, 0}}^2(0)- 
\frac{f_t^2}{2^{11}}\,      \frac{1 }{\epsilon}\,
\,q^4 R^2+  \frac{1}{R^2}\, \cO(q^2 R^2)
\end{eqnarray}
with $m_{\phi_{H,0}}^2(0)$ given by the first line in (\ref{div})
with replacements\footnote{The
series of $\Li_3$ in eq.(\ref{div}) due to $\cJ_2$'s in (\ref{j2finite})
can be integrated analytically, using \cite{berndt} (Corollary 5.8 for n=1).}
 (\ref{j2finite}). Note that the above divergence $q^4 R^2/\epsilon$
 due to  $\cJ_1$'s  is   of {\it similar form}   to that cancelled 
by supersymmetry in the difference $\cJ_2[0,0,c]-\cJ_2[1/2,1/2,c]$
in eq.(\ref{div})  (see the  terms in eqs.(\ref{j2finite})
proportional to $c^2/\epsilon~\sim~q^4 R^2/\epsilon$).
One notices the presence of a power four of the momentum scale in the
contribution in (\ref{dr}). 
The emergence of the divergence  in (\ref{dr})  shows  that higher 
(dimension) derivative operators 
are required  as one-loop counterterms for $m_{\phi_{H,0}}^2(q^2)$. 
The one-loop counterterm is\footnote{In (\ref{prt}), (\ref{prt2})
 the mass dimension 
of the superfields $H$, $H^c$ is  $3/2$.}
a ``brane'' N=1 interaction\footnote{One could also 
think of a N=2 ``bulk''   counterterm to generate  $q^4 R^2/\epsilon$
required in (\ref{dr})
\begin{eqnarray}\label{prt}
\int\! d^4 x \,dx_5 \int  d^2 \theta \, d^2\overline \theta
\,\, R^2 \, \Big( H^\dagger  \,\Box \, H
+ H^{c \dagger}  \,\Box \, H^c\Big) \sim \int d^4 x \,
 R^2 \, \phi_{H,0}^\dagger \, \Box^2 \, \phi_{H,0}+\cdots
\end{eqnarray}
but this is not possible since the $H^c$ dependent term
is not generated, ($H^c$  has no Yukawa coupling, 
see eq.(\ref{int})).}
\begin{equation}\label{prt2}
\int \! d^4 x \,dx_5 \!\int \! d^2 \theta \, d^2\overline \theta
\,\delta(x_5)\, \lambda^2 \, H^\dagger \Box  H  
\sim \! f_t^2\!
\int\! d^4 x \,
 R^2 \!\!\! \sum_{n,p\geq 0} \phi_{H,n}^\dagger \Box^2  \phi_{H,p}
\sim \! f_t^2\!
\int \!\! d^4 x \,
 R^2 \, \phi_{H,0}^\dagger  \Box^2  \phi_{H,0}
+...
\end{equation}
where we used that $\lambda^2=(\pi R)^3\,f_t^2$. This interaction
recovers in momentum space the one-loop correction in eq.(\ref{dr}).
If the compactification scale $1/R$ is high the suppression 
of the counterterm  is important, below this scale the model is 
essentially four-dimensional and 
$m_{\phi_{H, 0}}^2(q^2\!\ll\! 1/R^2)\!
\approx\! m_{\phi_{H, 0}}^2(0)\!\sim\! 1/R^2$.
However,  this behaviour   changes significantly for  models 
with  a ``large'' radius  $R$ or when $q^2 R^2\!\geq\!1$.
The presence of a  higher derivative 
operator  with an arbitrary (finite) coefficient  is relevant 
for physics at   $q^2\!\geq\! 1/R^2$ and can  be regarded as an effect of the
non-renormalisability of the initial 5D  theory. It can also 
affect the predictive power of  such models. Finally, 
$m^2_{\phi_{H,0}}(q^{\prime 2})- m^2_{\phi_{H,0}}(q^2)$ can easily  be
evaluated using $\cJ_{1,2}$ of eqs.(\ref{j1finite}), (\ref{j2finite}) 
or their generalisations (\ref{j1general}), (\ref{j2general}), to
provide the behaviour of the (mass)$^2$ under UV  scaling 
of the momenta, with applications to  phenomenology.

The pole in (\ref{dr}) due to $\cJ_1$ is just a result of
compactification: it  would  not be
present if we included  only a fixed, finite number of levels,
even if this were done such as to respect their (N=2) multiplet
structure.  To see this consider the situation when 
the sums over the Kaluza-Klein levels which enter $\cJ_1[1/2,1/2,c]$
and  $\cJ_1[0,0,c]$  are restricted 
to a finite number of modes, hereafter called $s_i$ and $s_i'$ 
with $i=1,2$.
The equivalent of $\cJ_1$ defined in (\ref{jjjs})  and present in
 (\ref{b}), (\ref{f}) is then

\begin{eqnarray}\label{eq14}
\!\!\!\!\cJ_1^*\,[\frac{1}{2},\frac{1}{2},c]\!\! &\!\equiv\! &\!\! 4\!\!
\!\sum_{0\leq n_{i}\leq s_i}
\int_0^\infty \!\!\frac{dt}{t^{1-\epsilon/2}}
\,e^{-\pi t\, [\,a_1 (n_1+1/2)^2 +a_2 (n_2+1/2)^2 +c]}
= \frac{2}{\epsilon}\, ( 4\, s_1 s_2) +\cO(\epsilon^0)\qquad\qquad
\nonumber\\
\nonumber\\
\!\!\!\!\!\!\cJ_1^*[0,0,c]\!\! &\!\equiv &\!\! 4\!\!
\!\sum_{0\leq n_{i}\leq s_i'}\frac{1}{2^{\delta_{n_1,0}}}
\frac{1}{2^{\delta_{n_2,0}}}
\int_0^\infty\! \frac{dt}{t^{1-\epsilon/2}}
\,e^{-\pi t\, [ \,a_1 n_1^2 +a_2 n_2^2 +c]}
= \frac{2}{\epsilon} \,(2 s_1'-\! 1)(2 s_2'-\! 1)\! +\!\cO(\epsilon^0)\quad
\\
\nonumber
\end{eqnarray}
Therefore  if the KK towers were restricted 
to a finite level the divergences in $\cJ_1^*$ are 
independent of $c\sim q^2 R^2$ which was not the case of $\cJ_1$ in 
eq.(\ref{j1j2}).  This shows that the limits 
$s_i\!\rightarrow\! \infty$ (or $s_i'\!\rightarrow\! \infty$)
 and $\epsilon \!\rightarrow\! 0$ 
of $\cJ_1^*$  do not commute and lead to different 
divergences\footnote{though their finite part is  the
same in the continuum limit, when 
an extra  divergence in $s_i$ also emerges in (\ref{eq14}).}.  From
eqs.(\ref{gre}), (\ref{b}), (\ref{f}), (\ref{jjjs}), (\ref{eq14})
we find the contribution of $\cJ_1^*$ to   $m_{\phi_{H,0}}^2(q^2)$

\begin{eqnarray}\label{divf}
- m_{\phi_{H,0}}^2(q^2)\bigg\vert_{\cJ_1^*}\!\!
& = & \!\!
\frac{1}{4} \frac{f_t^2}{(4\pi R)^2} \,(q^2 R^2)\!
\int_0^1  dx  \bigg[
x(x- 1) \cJ_1^*[0,0,c]
-(1-x)^2 \cJ_1^*[1/2,1/2,c]
+\cO(\epsilon^0)\bigg]\nonumber\\
\nonumber\\
&=& - \frac{f_t^2}{192 \,\pi^2}\, 
\bigg[\frac{q^2}{\epsilon} \bigg] 
\Big( (2s_1'-1)(2s_2'-1)+8 s_1 s_2\Big) \, +\cO(\epsilon^0)
\\
\nonumber
\end{eqnarray}
The divergence $q^4 R^2/\epsilon$ in eq.(\ref{dr})  has changed 
into $q^2/\epsilon$, eq.(\ref{divf}) in the case of including only a 
finite number of modes in each $\cJ_1$. The latter divergence is the expected
wavefunction renormalisation of the chiral superfield of 
the zero-mode  scalar. Therefore the presence of higher 
dimension (derivative)  operators responsible for $q^4 R^2/\epsilon$ 
is  a consequence  of summing over infinitely many
modes\footnote{An observation is in place here: one
may argue  that in general  radiative corrections from 
massive modes ``winding''  around compact dimension(s) can only 
induce finite loop corrections.  This  argument ignores the   
degeneracy of the modes and thus is not always correct
(this is the case where we have {\it two} (infinite) Kaluza-Klein  sums).}
(or ``large enough'' a number) 
and thus of the compactification of the initial 5D theory.

\section{Further Remarks and Conclusions.}

In  a minimal  5D N=1 supersymmetric model compactified 
on $S_1/(Z_2\!\times\! Z_2')$ with a brane-localised superpotential
 it was shown that  higher derivative 
operators  emerge  as one-loop counterterms 
to the mass of the zero-mode scalar field.
This finding is relevant for  more realistic  orbifold models  
\cite{dpq}-\cite{mr},
where this scalar field is  usually identified with the Higgs boson,
and our technical results can easily be used in such models.
It was shown that the emergence of such operators 
as counterterms is a direct effect of compactification, i.e. 
of  the integration of  infinitely many modes  of  the initial 
theory. The presence of such counterterms  is not surprising 
in the end, if we remember that the 5D theory  although 
supersymmetric, is nevertheless  non-renormalisable, but 
such effect was so far overlooked.
The result was obtained using a DR regularisation 
scheme for the 4D momentum integrals.
While  suppressed at low scales $q^2\!\ll\! 1/R^2$,
 where the theory is
essentially four dimensional, the operators 
 play an increasingly important role as we
scale the momentum $q^2$  to UV values $q^2\!\gg\! 1/R^2$. 
Our results  can also be used to investigate  $m_{\phi_{H,0}}^2(q^2)$ 
behaviour at $q^2\!\sim\! 1/R^2$ and under the UV scaling of $q^2$. 
Such operators  may also appear  in a 
similar way in the  one-loop corrected  Yukawa coupling.
The situation is  similar to the case of  radiative 
corrections to the gauge couplings in models with {\it two} 
compact dimensions, where higher derivative operators emerge
as one-loop counterterms in a very similar way~\cite{Ghilencea:2003xj}.

A natural question  is whether such operators can be generated
for other one dimensional orbifolds. This depends on the orbifold type
and  the interaction considered.
The technical reason for their presence 
in $S_1/(Z_2\!\times\! Z_2')$ was the (degeneracy  of the levels due
to the) existence of {\it two} 
Kaluza-Klein sums in the one-loop two-point Green function.
This is in turn due to the  brane-localisation of  interaction  (\ref{int}) 
with  {\it all fields} living in the ``bulk''. In momentum  space, 
with external legs of the  diagrams
fixed to zero-level ($\phi_{H,0}$), this lead to two  Kaluza-Klein 
sums in front of the loop-integral of the correction to the
(mass)$^2$;  their evaluation generated the higher derivative  operators.
Similar arguments apply  if one considered the {\it same} brane-localised 
interaction  in  $S_1/Z_2$ orbifolds and then  
higher derivative operators may emerge as one-loop counterterms
to  the mass of the zero-mode scalar. 
 However, if   the brane-localised 
interaction  (\ref{int}) had  only {\it one} ``bulk'' field, 
with the other two genuine 4D ``brane'' fields, overall
 only {\it one} Kaluza-Klein 
sum would be  present in the correction to the mass of
a (brane) scalar field. In that case such operators are not
generated at one-loop, but at higher loops  they can
 be present. This last observation 
also applies to a bulk interaction rather 
than the brane-localised interaction discussed above.

There is another, simpler way to understand these statements, based on 
purely dimensional arguments applied to the localised interaction 
$\lambda\, Q\, U\, H$ of (\ref{int}) and the general form of 
the localised counterterm  $(\lambda^2)^n H^\dagger \Box H$
of (\ref{prt2}), where we used that the latter must not 
depend on $R$.  If all these fields are bulk fields, $\lambda$ has
mass  dimension $[\lambda]=-3/2$ giving $n=1$ i.e. the counterterm
appears at one-loop,  
in agreement with the findings of the paper and with eq.(\ref{prt2}). 
However, if the  interaction has two genuine brane
fields and one bulk  field, $[\lambda]=-1/2$. Then the 
local counterterm  giving $q^4$ dependence has, if $H$ is a  brane 
field, $n=2$, thus it is generated  at the two-loop level, as already
argued above. If $H$ is a bulk field instead, dimensional arguments give 
that  $n=3$, thus such counterterm arises at three-loop only.
Similar considerations can be made for the bulk interactions using that
the gauge coupling also has mass dimension $-1/2$. 
These observations can be used when building higher dimensional
models, to assess the importance and avoid the presence 
 of higher derivative operators
(counterterms) at  small number of loops.

Such operators  are relevant for phenomenology. 
The presence of higher derivative operators  with  
arbitrary  coefficients which are new parameters in the theory
(depending on  its UV completion) may
affect the predictive power   of the models, particularly 
 if  the compactification scale is small.  
This can be important for the 
phenomenology of 5D orbifold  models with  a large extra-dimension, 
although only a model-by-model   study  can provide a  definite answer 
to the importance of  the effects of such operators.

The above analysis only discussed the case of a ``local''
interaction  at one fixed point of the orbifold
$S_1/(Z_2\!\times\!Z_2')$.  However, the ``global'' structure 
of  the orbifold  can involve
additional interactions (at the other  fixed points),
not considered here. Such ``global'' structure  
can  only be seen by studying 
``overlapping''  interactions originating from  different 
fixed points, and that may reveal new effects.
 This involves  calculations beyond 
one-loop order and  is beyond the purpose of this work.

Finally, since field theory orbifolds are  now extensively  
used for model building,  we  provided in  Appendix~C  a
generalisation of the  method used in the text for computing series 
of  divergent loop-integrals of the  radiative corrections generic 
in orbifold compactifications. The method applies for a  more general 
Kaluza-Klein spectrum  than that used in the text\footnote{See 
for example the spectrum in  eq.(\ref{mass}) in the Appendix and
  compare against that in eq.(\ref{mss}).}.  
The approach pays special attention to regularisation subtleties
within a  DR scheme  for the  momentum  integrals, to show a far richer 
set of  divergences than that of  the specific example discussed 
above.  We hope this will help a careful investigation  of
more realistic models.

\vspace{1cm}
 \noindent
\section*{\bf Acknowledgements.}
The author thanks S. Groot Nibbelink, R. Horgan,  H.~P.~Nilles, 
F.~Quevedo and G. Servant for many discussions on related topics. 
This work  was supported by a postdoctoral research grant from  the Particle 
Physics and Astronomy Research Council (PPARC), U.K.

\newpage

%%%%%%%%%%%%%%%%%%%%%%%%%%%%%%%%%%%%%%%%%%%%%%%%%%%%%%%%%%

\section*{Appendix}
\def\theequation{\thesubsection-\arabic{equation}} 
\def\thesubsection{A} 
\subsection{One-loop Integrals in DR.}\label{appa}
Calculations in DR (see for example \cite{IZ}) give that 
  ($d=4-\epsilon$, $\epsilon\ra 0$)

\begin{eqnarray}
\cK_1\!\! & \equiv  &\!\! \int \frac{d^d p}{(2 \pi)^d}
\frac{p_\mu}{((p+q)^2+m_2^2)(p^2+m_1^2)}
 =  \frac{-q_\mu}{(4\pi)^{\frac{d}{2}}}
\int_0^1 d x \,x \, \Gamma\big[2-{d}/{2}\big]
\bigg[ L(x,q^2,m_{1,2})\bigg]^{\frac{d}{2}-2}
\nonumber\\
\nonumber\\
\nonumber\\
\cK_2\! \!& \equiv &\!\! \int 
\frac{d^d p}{(2 \pi)^d} \frac{1}{((p+q)^2+m_2^2)(p^2+m_1^2)}
 =\,  \frac{1}{(4\pi)^{\frac{d}{2}}}\,
\int_0^1 d x \, \,\, \Gamma\big[2-{d}/{2}\big]
\bigg[ L(x,q^2,m_{1,2})\bigg]^{\frac{d}{2}-2} 
\nonumber\\
 \nonumber\\
 \nonumber\\
\cK_3\!\! & \equiv &\!\!  \int \frac{d^d p}{(2 \pi)^d} 
\frac{p_\mu p_\nu}{((p+q)^2+m_2^2)(p^2+m_1^2)}
 =  \frac{1}{(4\pi)^{\frac{d}{2}}}\,\,
\frac{\delta_{\mu\nu}}{2}\,\,
\int_0^1 d x \,\, \,\,\Gamma\big[1-{d}/{2}\big] 
\bigg[ L(x,q^2,m_{1,2})\bigg]^{\frac{d}{2}-1} 
\nonumber\\
\nonumber\\
&& \hspace{5.5cm}
+\, \frac{1}{(4\pi)^{\frac{d}{2}}} \,\,q_\mu q_\nu\,\,
\! \! \int_0^1 d x \,  x^2\,
\Gamma\big[2-  {d}/{2}\big] 
\bigg[ L(x,q^2,m_{1,2})\bigg]^{\frac{d}{2}-2} 
\nonumber
\end{eqnarray}
where 
\begin{equation}
L(x,q^2,m_{1,2})\equiv x \,(1-x)\, q^2 +x\,  m_2^2 +(1-x)
m_1^2
\end{equation}

\vspace{0.3cm}
\noindent
 With the notation  $\cL \equiv R^2  L(x,q^2,m_{1,2})$ and 
 $\sum_\mu \delta_{\mu\mu}\!=\!d$ one finds:

\begin{eqnarray}
\cK_b &\equiv & \int \frac{d^d p}{(2 \pi)^d} 
\frac{(p+q)^2}{((p+q)^2+m_2^2)(p^2+m_1^2)}\nonumber\\
\nonumber\\
& = & \frac{(R^2)^{1-\frac{d}{2}}}{(4\pi)^{\frac{d}{2}}}\,\,
\int_0^1 d x \,
\bigg[
\frac{d}{2} \,\cL^{{d}/{2}-1}\, \Gamma[1-{d}/{2}]\,
 + q^2 R^2 \,  (1-x)^2\,\cL^{d/2-2}\, \Gamma[2-  {d}/{2}]\,\bigg]
\qquad\qquad\qquad
\label{j4} \\
\nonumber\\
\nonumber\\
\cK_f& \equiv & 
 \int \frac{d^d p}{(2 \pi)^d} 
\frac{p^2+p.q}{((p+q)^2+m_2^2)(p^2+m_1^2)}\nonumber\\
\nonumber\\
& = & \frac{(R^2)^{1-\frac{d}{2}}}{(4\pi)^{\frac{d}{2}}}\,\,
\int_0^1 d x \,\bigg[\frac{d}{2}\, \cL^{{d}/{2}-1}\,
\Gamma\big[1-{d}/{2}\big] + q^2 R^2\, x\,(x-1)\,\,\cL^{{d}/{2}-2}\,
\Gamma\big[2- {d}/{2}\big]  \bigg]
\label{j5}\\
\nonumber
\end{eqnarray} 
Using the identity $\Gamma[s]\, (\pi \cL)^{-s} =
\int_0^\infty dt \, t^{s-1} \, e^{-\pi\cL \, t}\,$
 one then finds eqs.(\ref{b}),(\ref{f}) in the text.

%%%%%%%%%%

 \newpage
\def\theequation{\thesubsection-\arabic{equation}} 
\def\thesubsection{B} 
\setcounter{equation}{0}
\subsection{Series of Divergent Integrals in DR.}\label{J1andJ2}
$\bullet$ Following calculations in \cite{Ghilencea:2003xy} 
(see also \cite{Ghilencea:2003kt}),
one shows  that  for $c, a_1, a_2 >0$ and real $c_1, c_2$

\begin{eqnarray}\label{j1general}
\!\!
\cJ_1[c_1,c_2,c] 
 \!& \equiv  &\!\!  \sum_{n_1,n_2\in\bZ}
\int_0^\infty \frac{dt}{t^{1-\epsilon/2}}
\, e^{-\pi \, t\, [\,c+a_1 (n_1+c_1)^2+a_2 (n_2+c_2)^2]}
\nonumber\\
\nonumber\\
 & =&  \!\!\! \frac{\pi c}{\sqrt{a_1 a_2}}
\bigg[\frac{-2}{\epsilon}\bigg]
+ 2\pi 
\bigg[\frac{a_1}{a_2}\bigg]^\frac{1}{2}
\bigg[\frac{1}{6}+\Delta_{c_1}^2
-\Big(\frac{c}{a_1}+\Delta_{c_1}^2\Big)^{\frac{1}{2}}
 \bigg]
\nonumber\\
\nonumber\\
&+&\frac{\pi c}{\sqrt{a_1 a_2}} \ln\Big[4\pi a_1
    e^{\gamma+\psi(\Delta_{c_1})+\psi(-\Delta_{c_1})}\Big] 
 -   \sum_{n_1\in\bZ}
\ln\Big\vert 1-e^{\frac{-2\pi}{\sqrt a_2} [c+a_1
    (n_1+c_1)^2]^\frac{1}{2}+ 2i\, \pi \,c_2}\Big\vert^2
\nonumber\\
\nonumber\\
& +&\!\! \bigg[\pi \frac{a_1}{a_2}\bigg]^\frac{1}{2} 
\!\sum_{p\geq 1}^\infty \frac{\Gamma[p+1/2]}{(p+1)!}
\Big[\frac{-c}{a_1}\Big]^{p+1} 
\Big[\zeta[2p+1,1+ \Delta_{c_1}]+\zeta[2p+1,1-\Delta_{c_1}]\Big]
\end{eqnarray}

\noindent
$\zeta[z,q]$ is the Hurwitz  
Zeta function,  $\psi(x)=d/dx \,\ln \Gamma[x]$. 
The result only depends on the  fractional part of
$c_i\,$ ($i=1,2$) defined by
$\Delta_{c_i} \equiv c_i-[c_i]$ with $0\!\leq\! \Delta_{c_i}\!<\! 1$ and 
$[c_i]\in \bZ$. The exclusion of a finite set of
modes/levels  (such as a zero mode, etc) from  the  two sums
in the definition of $\cJ_1$ would however bring in  a dependence 
of $\cJ_1$ on $c_i$ (rather than $\Delta_{c_i}$)  and also  
 additional poles in~$\epsilon$.
With $u\equiv \sqrt{a_1/a_2}$ one has
\begin{eqnarray}\label{prw}
\!\!\!\!\!
\cJ_1[c_1,c_2,c\ll\! 1]  & = &\!\!\!\!\!  \sum_{n_1,n_2\in\bZ}
\int_0^\infty \frac{dt}{t^{1-\epsilon/2}}
\, e^{-\pi \, t\, [\,c+a_1 (n_1+c_1)^2+a_2 (n_2+c_2)^2]}
\\
\nonumber\\
& =&  \!\!\! \frac{\pi c}{\sqrt{a_1 a_2}}
\bigg[\frac{-2}{\epsilon}\bigg]
-\ln 
\bigg\vert \frac{\vartheta_1 (c_2 -i u\, c_1\vert i u)}{(c_2 - i u
  c_1) \, \eta(i \,u)}
\,e^{-\pi  u\, c_1^2} \bigg \vert^2\!\!\! -\ln\! \Big[(
c+a_1 c_1^2 + a_2 c_2^2)/a_2\Big]
\nonumber
\end{eqnarray}
Above we only kept the term proportional to
$c/\epsilon$ because the limits $c\! \ra\! 0$ and $\epsilon \!\ra\! 0$ do not
commute and neglected $\cO(c)$ terms; 
we used 
\begin{eqnarray}
\vartheta_1(z\vert \tau) \equiv   -i \sum_{n\in\bZ}
(-1)^n \, e^{i \pi \tau (n+1/2)^2}\, e^{(2 n+1) i\pi z},\qquad 
\eta(\tau) \equiv   e^{i\pi \tau/12} \prod_{n\geq 1} (1-e^{2 i\pi \tau n})
\end{eqnarray}
for the Jacobi Theta function $\vartheta_1$ 
and Dedekind  function $\eta$, respectively.

For the particular cases encountered in the text, eq.(\ref{prw}) gives

\begin{eqnarray}\label{eqb4}
\cJ_1[0,0,c\ll 1]
& = & 
\frac{\pi c}{\sqrt{a_1 a_2}}
\bigg[\frac{-2}{\epsilon}\bigg]
-\ln \bigg[4 \pi e^{-\gamma} \big\vert
 \eta (i\, u)\big\vert^4 \frac{1}{a_2}\bigg]
-\ln (\pi e^\gamma c)
\nonumber\\
\nonumber\\
\cJ_1[\frac{1}{2},\frac{1}{2}, c\ll 1]& = & 
\frac{\pi c}{\sqrt{a_1 a_2}}
\bigg[\frac{-2}{\epsilon}\bigg]
-\ln  \bigg\vert \frac{\vartheta_1 
(1/2 -i u/2\vert i u)}{\eta(i \,u)} \bigg \vert^2
+\frac{\pi}{2} u,\quad u\equiv \sqrt{a_1/a_2}
\end{eqnarray}
used in eq.(\ref{j1finite}).

%%%%%%%%%%%%%%%%%%%%%%%%%%%%%%%%%%%%%%%%%%%%%%%%%%%%%%%%%%%%%%%%

\newpage
\noindent
$\bullet$ 
A similar calculation gives that,  for $c, a_1, a_2>0$ and $c_1, c_2$ real
\begin{eqnarray}\label{j2general}
\cJ_2[c_1,c_2,c]\!\!\! & \equiv  &  \sum_{n_1,n_2\in\bZ}
\int_0^\infty \frac{dt}{t^{2-\epsilon/2}}\,
\, e^{-\pi \, t\, [\,c+a_1 (n_1+c_1)^2+a_2 (n_2+c_2)^2]}
\nonumber\\
\nonumber\\
&=&\!
-\frac{\pi^2 c^2}{2 \sqrt{a_1 a_2}}\bigg[\frac{-2}{\epsilon}\bigg]
+\!\frac{\pi^2 a_1^{3/2}}{3 a_2^{1/2}} 
\bigg[\frac{1}{15}\! - 2 \Delta_{c_1}^2 
\Big(
1 + \!\Delta_{c_1}^2 +  \frac{3 \,c}{a_1}\Big)\! - \! \frac{c}{a_1} 
+ 4 \Big(\frac{c}{a_1}+\Delta_{c_1}^2\Big)^{\frac{3}{2}}\bigg]
\nonumber\\
\nonumber\\
&+&
\sum_{n_1\in\bZ} 
\big[ a_2 (c+a_1 (n_1+c_1)^2)\big]^{\frac{1}{2}}
 \Li_2\Big(e^{-2\pi [ (c/a_2+a_1 (n_1+c_1)^2/a_2)^{\frac{1}{2}}-i \, 
c_2]}\Big)
 \nonumber\\
 \nonumber\\
 &+ &
\frac{a_2}{2\pi}\sum_{n_1\in\bZ} \Li_3\Big(e^{-2\pi [  (c/a_2+a_1
  (n_1+c_1)^2/a_2)^{\frac{1}{2}} -i c_2]}\Big)
-
\frac{\pi^2 c^2}{2 \sqrt{a_1 a_2}} \ln \Big[4\pi\,
  e^{\gamma+\psi(\Delta_{c_1})+\psi(-\Delta_{c_1})}\Big]
\nonumber\\
\nonumber\\
&+&
\frac{\pi^{3/2} c^2}{\sqrt {a_1 a_2}}
\sum_{p\geq 1}^{\infty} \frac{\Gamma[p+1/2]}{(p+2)!}
\Big[\frac{-c}{a_1}\Big]^p 
\Big[\zeta[2p+1,1+\Delta_{c_1}]+
\zeta[2p+1,1-\Delta_{c_1}]\Big]+c.c.\label{rtg}
\end{eqnarray}

\noindent
where ``c.c'' only applies to the PolyLogarithm functions Li$_2$ and Li$_3$.
Similar to the previous case, 
we introduced the fractional part of $c_i$ defined as 
 $\Delta_{c_i}\equiv c_i-[c_i]$, 
$0\!\leq\! \Delta_{c_i}\!<\! 1$ and $[c_i]\in \bZ$.
 
Eq.(\ref{rtg}) gives:
\begin{eqnarray}\label{ptw}
\!\!\!\!\!\!
\cJ_2[c_1,c_2,c\ll 1] & = &  \sum_{n_1,n_2\in\bZ}
\int_0^\infty \frac{dt}{t^{2-\epsilon/2}}\,
\, e^{-\pi \, t\, [\,c+a_1 (n_1+c_1)^2+a_2 (n_2+c_2)^2]}
\nonumber\\
\nonumber\\
&=& 
-\frac{\pi^2 c^2}{2 \sqrt{a_1 a_2}}\bigg[\frac{-2}{\epsilon}\bigg]
+\frac{\pi^2 a_1}{3}\bigg[
\frac{a_1}{a_2}\bigg]^\frac{1}{2} \bigg[\frac{1}{15}-2 \Delta_{c_1}^2 (1-
 \Delta_{c_1})^2 \bigg]
\nonumber\\
\nonumber\\
&+&\bigg[
\sqrt{a_1 a_2} \sum_{n\in\bZ} \vert n+c_1 \vert\, \Li_2(e^{-2\pi i z})
+\frac{a_2}{2\pi}\sum_{n\in\bZ} \Li_3(e^{-2\pi i z})+c.c.\bigg]\qquad
\end{eqnarray}

\noindent
with $c, a_1, a_2>0$  and  $z=c_2-i \, (a_1/a_2)^\frac{1}{2}
\vert n+c_1\vert$. We kept the term proportional to
$c^2/\epsilon$ because the limits $c\ra 0$ and 
$\epsilon \ra 0$ do not commute.
Eq.(\ref{ptw}) finally gives

\begin{eqnarray}\label{jj1jj2}
\cJ_2[0,0,c\!\ll \!1] \!\!\!\!& = &\!\!\!\!\!
\frac{-\pi^2 c^2}{2 \sqrt{a_1 a_2}}\bigg[\frac{-2}{\epsilon}\bigg]
\!+\! \frac{\pi^2 a_1 u}{45}
\!+\! \frac{a_2}{\pi} \sum_{n\in\bZ}\!\bigg[ 
\Li_3(e^{-2 \pi\, u\,\vert n \vert})
\!+\! 2\pi u  \vert n\vert \,
 \Li_2(e^{-2\pi u \vert  n\vert})\bigg]\qquad\qquad\quad
\end{eqnarray}
\begin{eqnarray}
\cJ_2 [\frac{1}{2},\frac{1}{2},c\!\ll \! 1 ] \!\!\!\! & = & 
\!\!\!\!\!
\frac{-\pi^2 c^2}{2 \sqrt{a_1 a_2}}\bigg[\frac{-2}{\epsilon}\bigg]
\! -\! \frac{7 \pi^2 a_1 u}{360}  
\! +\!
\frac{a_2}{\pi}\!\sum_{n\in\bZ} 
\bigg[\Li_3(-e^{-2\pi u \,\vert n\! +\!\frac{1}{2}\vert})
\! + \! 2\pi u  \Big\vert n\! +\!\! \frac{1}{2} \Big\vert
\, \Li_2(-e^{-2\pi \,u \vert n+\frac{1}{2}\vert})\!\bigg]
\nonumber
% \\
% \nonumber
\end{eqnarray}
with
$$u\equiv \sqrt{{a_1}/{a_2}}$$
The two $\cJ_2$ in (\ref{jj1jj2}) were used in eq.(\ref{j2finite}).

\newpage
\def\theequation{\thesubsection-\arabic{equation}} 
\def\thesubsection{C} 
\setcounter{equation}{0}
\subsection{General evaluation of series of (divergent)
  loop-integrals using  DR.}
\label{appc}
When computing radiative corrections in orbifold compactifications 
one has to sum infinitely many  contributions from divergent loop
integrals (of 4D Feynman diagrams), of general structure  shown in 
eq.(\ref{gr}).  Here we show a careful evaluation of this expression
which generalises that encountered in the text eq.(\ref{gre}) with 
(\ref{mss}).
With minimal changes this can be applied to  one-loop calculations in 
 $S_1/(Z_2\!\times\! Z_2')$, in some $S_1/Z_2$ 
models at one-loop\footnote{Our regularisation can  be used in
 models where  series like (\ref{gr}) 
appear  at two-loop only  \cite{Delgado:2001xr}, eq.(4.39).} 
\cite{Arkani-Hamed:2001mi} and also in  two dimensional
compactifications,  given the presence of a double KK sum in 
(\ref{gr}). The technique can also be used in cases with 
one (rather than two) infinite sums.
The method is useful  for an easy evaluation of 
the overall divergence of the final result\footnote{See also 
\cite{GrootNibbelink:2001bx} for a related calculation.}.
Therefore we shall  compute
\begin{eqnarray}\label{gr}
\cH\equiv\sum_{k_1\geq 0} \sum_{k_2\geq 0}
\int \frac{d^d p}{(2\pi)^d}
\frac{\alpha \, p^2+\beta\, (p.q)+\gamma \, q^2+\delta}{ 
[ (q+p)^2+m_{k_2}^2 ]^n \, [p^2+m_{k_1}^2]^m}
\end{eqnarray}
Here $q$ is an  (Euclidean) external momenta. 
The loop integrals are regularised in DR with $d=4-\epsilon$,
$\epsilon\ra 0$. No additional regulator is
needed for the infinite sum(s).  $m_{k_{1,2}}$ are  KK masses in
the loops, and the sums are restricted to positive modes only due to
(orbifold) parity constraints.  
 $\alpha, \beta, \gamma$ are arbitrary, dimensionless parameters 
while $\delta$ has  dimension of (mass)$^2$. Here we take $n,m \geq 1$ 
and the most general structure for the  masses of the KK states
\begin{eqnarray}\label{mass}
m_{k_1}=\frac{u}{R}\,(k_1+c_1),\quad
m_{k_2}=\frac{1}{R}\,(k_2+c_2).
\end{eqnarray}
For generality we  introduced 
$u$ an arbitrary positive constant (dimensionless).
We denote $\cH_{k_1, k_2}$ the momentum integral in (\ref{gr}).
Standard integration techniques give  \cite{IZ}
\begin{eqnarray}\label{lpk}
\cH_{k_1, k_2}\!\!\!\! & = &\!\!\! 
\frac{1}{(4\pi)^{\frac{d}{2}}}
\frac{1}{\Gamma[n]\Gamma[m]}
\bigg\{\! \int_0^1\!\! dx\, x^{n-1}\, (1-x)^{m-1}\! 
\nonumber\\
\nonumber\\
&\times &\!\!\! \!\!\! 
\Big[ \Big(\delta  +q^2 (\gamma -x \beta +x^2 \alpha) \Big)\,
\Gamma[n+m-{d}/{2}] \,L^{\frac{d}{2}-n-m}
+ \alpha \,\frac{d}{2}\, \Gamma[n+m-{d}/{2}-1] \, 
L^{\frac{d}{2}-n-m+1}\Big]\bigg\}
\nonumber\\
\nonumber\\
L& \equiv & \, x(1-x)\, q^2 +(1-x)\, m_{k_1}^2+ x\, m_{k_2}^2
\quad\,
\end{eqnarray}
UV divergences in $\cH$ arise from the two Gamma functions in the 
square brackets above and possibly from the integrals over $x$. 
With $d=\!4\!-\!\epsilon$ poles in $\epsilon$ 
are present only if $n+m\!\leq\! 3$. To find the overall divergence of $\cH$ 
one  must evaluate the  infinite sums below, eq.(\ref{ifs})
with $s=\epsilon/2$; $\epsilon/2-1$ to order
$\cO(\epsilon^0)$. To find  the finite part of $\cH$ one further needs
the  $\cO(\epsilon)$  terms. We restrict the
calculation  to $\cO(\epsilon^0)$ only, to find the divergent part of $\cH$, 
but the analysis is easily extended to find its 
finite part as well. From (\ref{gr}), (\ref{lpk}) with
(\ref{mass}) we conclude that we must evaluate
\begin{eqnarray}\label{ifs}
\!\!\sum_{k_1\geq 0}\sum_{k_2\geq 0} 
\Big[c + \sigma_1 \,(k_1+c_1)^2 + \sigma_2\,(k_2+c_2)^2 \Big]^{-s}\!,
\quad
\sigma_1\!\equiv u^2 (1-x),\, \sigma_2\!\equiv x,\,\, c\!\equiv x(1-x) q^2 R^2
\end{eqnarray}
for $s=\epsilon/2$ and $-1+\epsilon/2$ at $\cO(\epsilon^0)$.
The evaluation of (\ref{ifs}) is known in the context of Epstein-like
functions. Such sums are usually evaluated for the whole set $\bZ$ of
integers. However, in orbifold constructions parity constraints 
require one perform the sums for positive integers only, and with
arbitrary $c_{1,2}$,  which is a more difficult task. 
To evaluate (\ref{ifs})  we rely on the general result for  
a one dimensional sum extensively studied in the literature 
\cite{elizalde},  
and then use the result in eq.(\ref{ifs}).
We introduce
\begin{eqnarray}\label{sume1}
E_1[c; s; \sigma_1, c_1] & \equiv &  \sum_{n_1\geq 0} [\sigma_1
  (n_1+c_1)^2+c]^{-s}
\end{eqnarray}
This has the  {\it asymptotic}  expansion (see \cite{elizalde}
for a proof)

\begin{eqnarray}\label{ff1}
E_1[c; s; \sigma_1, c_1] & \approx & {c^{-s}}\sum_{m\geq 0}
\frac{\Gamma[s+m]}{m!\, \Gamma[s]} 
\bigg[\frac{-\sigma_1}{c}\bigg]^m \zeta[-2 m, c_1]
+\frac{c^{1/2-s}}{2}\bigg[\frac{\pi}{\sigma_1}\bigg]^{\frac{1}{2}}
\frac{\Gamma[s-1/2]}{\Gamma[s]}
\nonumber\\
\nonumber\\
& + &
\frac{2 \pi^s}{\Gamma[s]}\cos(2\pi c_1) \sigma_1^{-\frac{s}{2}-\frac{1}{4}}
c^{-\frac{s}{2}+\frac{1}{4}}
\sum_{n_1\geq 1} n_1^{s-\frac{1}{2}}K_{s-\frac{1}{2}}\bigg(2\pi n_1 
\big({c}/{\sigma_1}\big)^{\frac{1}{2}}\bigg)\\
\nonumber
\end{eqnarray}
where $\zeta[q,a], a\not=0,-1,-2, \cdots$ is the Hurwitz Zeta function, 
$\zeta[q,a]=\sum_{n\geq 0} (n+a)^{-q}$ for $\textrm{Re}(q)>1$. 
$K$ is the usual modified Bessel function \cite{gr}.
The  singularities of  $E_1$ can arise for 
specific values of $s$,  from poles of the Gamma functions in the rhs 
of  the first line in\footnote{An alternative to 
using (\ref{ff1}) 
is to make a simple {\it binomial} expansion of the parenthesis under the
sum in (\ref{sume1}) in powers of $c/\sigma_1 \leq 1$ and  then just use the
definition of Hurwitz $\zeta$-function. 
 This easily gives:
\begin{eqnarray}\label{ff2}
E_1[c; s; \sigma_1, c_1] &  = & 
 \sigma_1^{-s}\sum_{k\geq 0} \frac{\Gamma[k+s]}{k! \,
  \Gamma[s]}\bigg[\frac{-c}{\sigma_1}\bigg]^k \, \zeta[2k+2s, c_1]
\end{eqnarray}
The singularities of $E_1$  given in (\ref{exp1}) 
arise now from either the Gamma functions (if $s$ is
a negative  integer or zero) and from the usual singularity of 
$\zeta$-function
at $2 k+2 s=1$,  (if $s$ is $\pm 1/2, -3/2, -5/2, -7/2,  \cdots$).}
eq.(\ref{ff1}). 
Using either eq.(\ref{ff1}) or (\ref{ff2}) one can easily show that
\begin{eqnarray}\label{exp1}
E_1[c; -\epsilon; \sigma_1, c_1] 
& = & 
\zeta[0,c_1]+\cO(\epsilon)
  \nonumber\\
\nonumber\\
E_1[c; -1-\epsilon, \sigma_1, c_1] & = & 
\sigma_1 \, \zeta[-2, c_1]+c \, \zeta[0,c_1]+\cO(\epsilon)
\nonumber\\
\nonumber\\
E_1[c; -\epsilon-{1}/{2}; \sigma_1, c_1]
& = & -\frac{c}{4\sqrt \sigma_1}\frac{1}{\epsilon}+\cO(\epsilon^0)
\nonumber\\
\nonumber\\
E_1[c; -\epsilon-{3}/{2}; \sigma_1, c_1]
& = & -\frac{3 c^2}{16  \sqrt \sigma_1}\frac{1}{\epsilon}+\cO(\epsilon^0)
\end{eqnarray}
where one can further use that $\zeta[0,x]=1/2-x$. 
These results will be used shortly to evaluate the double sums in~(\ref{ifs}).

\noindent
Further, eq.(\ref{ff1}) gives, 
after the replacement $c\!\ra\! c+\! \sigma_2 (n_2\!+\! c_2)^2$  and 
a summation   over $n_2$:
\begin{eqnarray}\label{exp2}
E_2[c; s; \sigma_1, \sigma_2; c_1, c_2]\!\!\!
& \equiv & \!\!\sum_{n_1\geq 0}\sum_{n_2\geq 0}
\bigg[\sigma_1(n_1+c_1)^2+\sigma_2 (n_2+c_2)^2+c\bigg]^{-s}  
\nonumber\\
\nonumber\\
& \approx &
\frac{\sigma_2^{-s}}{\Gamma[s]} \sum_{m\geq 0} \frac{(-1)^m}{m!} \Gamma[s+m]
\bigg[\frac{\sigma_1}{\sigma_2}\bigg]^m
\zeta[-2 m, c_1] \, E_1[c/\sigma_2; s+m; 1,  c_2]
\nonumber\\
\nonumber\\
& + &
\!\! {\sigma_2^{\frac12-s}}
\bigg[\frac{\pi}{\sigma_1}\bigg]^{\frac{1}{2}}\frac{\Gamma[s\! - \! 1/2]}{
  2 \,\Gamma[s]}
\, E_1[c/\sigma_2; s\! -\! 1/2; 1, c_2]\!
 + 
\frac{2\pi^s}{\Gamma[s]}\cos(2\pi c_1)\, \sigma_1^{-\frac{s}{2}-\frac{1}{4} }
\nonumber\\
\nonumber\\
&\times & \!\!\!\!\!\!
\sum_{n_1\geq 1} \sum_{n_2\geq 0} n_1^{s-\frac{1}{2}}
\big[\sigma_2(n_2+c_2)^2+c\big]^{-\frac{s}{2}+\frac{1}{4}}
K_{s-\frac{1}{2}}\bigg(\frac{2\pi n_1}{\sigma_1^{1/2}}
[\sigma_2(n_2+c_2)^2+c]^{\frac{1}{2}}\bigg)\qquad\,
\\
\nonumber
\end{eqnarray}
For an extended study of the properties of $E_2$ see~\cite{elizalde}.

 Using  now  eqs.(\ref{exp1}) in (\ref{exp2}), one finally 
finds from (\ref{exp2}) with $s=\epsilon/2$ or $s=-1+\epsilon/2$:

\begin{eqnarray}\label{rp1}
\cR_0 & \equiv &  \!\!
\sum_{n_1> 0}\sum_{n_2 > 0} 
\bigg[  \sigma_1 (n_1+c_1)^2 \!+\! \sigma_2 (n_2+c_2)^2\!+\!c
\bigg]^{-{\epsilon}/{2}}
\!\!\!\! = \!
- \frac{\pi}{4} \frac{c}{\sqrt{\sigma_1 \sigma_2}}
 + \zeta[0,1\!+\!c_1]\, \zeta[0,1\!+ \! c_2]\!+\!\cO(\epsilon)
\nonumber\\
\nonumber\\
\cR_1  &\!\! \equiv &\!\! 
\sum_{n_1 > 0}\sum_{n_2 > 0} 
\bigg[  \sigma_1 (n_1+c_1)^2 + \sigma_2 (n_2+c_2)^2 +c 
\bigg]^{1-\epsilon/2}\label{rp2}
\nonumber\\
\nonumber\\
& = & 
- \frac{\pi}{8}  \frac{c^2}{\sqrt{\sigma_1 \sigma_2}}
 + \frac{1}{3} \left(c_1+\frac{1}{2}\right) \!
\left(c_2+\frac{1}{2}\right) 
[\sigma_1 c_1 (c_1+1)+\sigma_2 c_2 (c_2+1) + 3 c ]
+\!\cO(\epsilon)\qquad\qquad\quad
\\
\nonumber
\end{eqnarray}

\noindent
which computes (\ref{ifs}). 
From (\ref{gr}),(\ref{mass}), (\ref{lpk}), (\ref{rp1}) we finally find
 the overall divergence of~$\cH$  

\begin{eqnarray}\label{klgh}
\cH \,\bigg\vert_{n=m=1}\!\!\!\!\!\! &= & \sum_{k_1\geq 0} \sum_{k_2\geq 0}
\int \frac{d^d p}{(2\pi)^d}
\frac{\alpha \, p^2+\beta\, (p.q)+\gamma \, q^2+\delta}{ 
[ (q+p)^2+ (k_2+c_2)^2/R^2 \,]\, \, [\,p^2+ (k_1+c_1)^2 u^2/R^2\,]}
\nonumber\\
\nonumber\\
 & = & \!\! \frac{1}{(4\pi)^2}\, \frac{2}{\epsilon}\;
\bigg[\,\delta 
-
\frac{\alpha}{3 \,R^2} \Big( u^2\, (c_1-1)
    c_1+(c_2-1) c_2\Big) + q^2 (\gamma-\beta/2)\bigg]
\Big(c_1-\frac{1}{2}\Big)\Big(c_2-\frac{1}{2}\Big)
\nonumber\\
\nonumber\\
& + & \frac{1}{(4\pi)^2}\, \frac{2}{\epsilon}\;\bigg[ -
\frac{\pi^2}{32 \, u} \, \delta\, q^2 R^2 
-\frac{\pi^2}{2^9 \, u}\, (2\, \alpha-8\, \beta +16\,\gamma\,)\,\,
q^4 R^2\bigg] +\cO(\epsilon^0), \\
\nonumber
\nonumber\\
\nonumber\\
\cH \,\bigg\vert_{n+m=3}\!\!\!\!\!\
 & = &  -\frac{1}{(4\pi)^2}\, \frac{2 }{\epsilon}\;\alpha\,
\bigg[\frac{\pi^2}{32\, u} q^2 R^2 -\Big(c_1-\frac{1}{2}\Big)
\Big(c_2-\frac{1}{2}\Big)\bigg]+\cO(\epsilon^0)\label{rpd}
\end{eqnarray}
for arbitrary $\alpha,\beta,\gamma,\delta, c_1, c_2$ and $u> 0$.
In (\ref{rpd}) we used that $n\geq 1, m\geq 1$.

\noindent
Note also a generalisation of the fermionic contribution presented in
the text
\begin{eqnarray}\label{thr}
\cF &\equiv  & 
\sum_{k_1\geq 0} \sum_{k_2\geq 0}
\frac{1}{2^{\delta_{k_1,0}}}
\frac{1}{2^{\delta_{k_2,0}}}
\int \frac{d^d p}{(2\pi)^d}
\frac{\alpha \, p^2+\beta\, (p.q)+\gamma \, q^2+\delta}{ 
[ (q+p)^2+(k_2+c_2)^2/R^2\, ] \,\, [\,p^2+ (k_1+c_1)^2\, u^2/R^2 ]}
\nonumber\\
\nonumber\\
&=&\frac{1}{(4 \pi)^2}\,\frac{2}{\epsilon}\,
\bigg[\delta-\frac{\alpha}{6\,R^2} \Big(1+u^2+2\, 
(c_1^2 u^2+c_2^2)\Big) +q^2(\gamma-\beta/2)\bigg] c_1 c_2
 \nonumber\\
\nonumber\\
&+& 
\frac{1}{(4\pi)^2}\, \frac{2}{\epsilon}\,
\bigg[-\frac{\pi^2}{32 u}\, \delta\, q^2 R^2 -\frac{\pi^2}{2^9 u}
  \,(2\,\alpha-8\beta+16\gamma)\, q^4 R^2\bigg]+\cO(\epsilon^0)\\
\nonumber
\end{eqnarray}
which can be compared against (\ref{klgh}).

In conclusion $\cH$ of (\ref{klgh}) has divergences which are both
dependent and independent of $q^2$. The later  are not all ``seen'' 
in the  special cases $c_1$ or $c_2$ is equal to $1/2$ or 0 or when  
$\alpha=\delta=0$ (similar considerations apply to $\cF$ of (\ref{thr})).
The $q^2$ or $q^4$-dependent divergences are associated with
wavefunction renormalisation or higher dimension derivative operators 
and are absent only for special values of the parameters (e.g. $\delta=0$,
$\gamma\!=\!\beta/2$ and respectively $2\alpha-8\beta+16\gamma=0$).
One can also compute the $\cO(\epsilon^0)$ terms in (\ref{klgh})
following the same approach. Note that in some cases additional 
singularities in $x$ (not in $\epsilon$) may appear \cite{IZ} from 
$\cO(\epsilon^0)$  terms,  from  the integrals over~$x$.
The above results can be used in one and two dimensional
compactifications while summing infinitely many Feynman diagrams, and
provide a careful approach to computing their overall divergence.

As an application to the $S_1/(Z_2\times Z_2')$ orbifold,  eq.(\ref{klgh})
with $u=1$, $\delta=0$, $\gamma=1$, $\beta=2$, $\alpha=1$, $c_1=c_2=1/2$,
$R\ra R/2$ gives, up to an overall sign, 
the contribution for bosons encountered  in the
text, eqs.(\ref{b}). Similarly, eq.(\ref{thr}) for 
$u=1$, $\delta=0$, $\gamma=0$, $\beta=1$, $\alpha=1$, $c_1=c_2=0$,
$R\ra R/2$ gives  the fermionic contribution in the text, eq.(\ref{f}).
Note that for $c_i\not=0,1/2$  the sums in $\cH$,
$\cF$ over positive integers cannot be written in terms of 
sums over the  whole set $\bZ$,  as  done in the text
for their counterparts in eqs.(\ref{b}), (\ref{f}) with
$c_i=0,1/2$. In such case the results in Appendix~\ref{J1andJ2} (with 
Appendix~\ref{appa})  which evaluate the corresponding sums over the 
whole set $\bZ$ cannot be used  anymore and  one has to apply  
instead the results of this section.

%\newpage

\end{document}